\documentclass[runningheads,a4paper]{llncs}
\usepackage[width=122mm,left=44mm,paperwidth=210mm,height=193mm,top=52mm,paperheight=297mm]{geometry}

\usepackage{amssymb}
\usepackage{amsmath}
\usepackage{graphicx}
\graphicspath{{./figures/}}

\usepackage{url}
\usepackage{setspace}
 \normalsize

\begin{document}
\sloppy

\title{A Measurement Study of TCP Performance for Chunk Delivery in DASH}

\author{Wen Hu\inst{1}%
\and Zhi Wang\inst{2} \and Lifeng Sun\inst{1}}

\institute{Tsinghua National Laboratory for Information Science and Technology\\ Department of Computer Science and Technology, Tsinghua University 
\and
Graduate School at Shenzhen, Tsinghua University\\
\email{\{hu-w12@mails., wangzhi@sz., sunlf@\}tsinghua.edu.cn}}

\maketitle

\begin{abstract}

Dynamic Adaptive Streaming over HTTP (DASH) has emerged as an increasingly popular paradigm for video streaming~\cite{baochun-tomccap-streaming2013}, in which a video is segmented into many chunks delivered to users by HTTP request/response over Transmission Control Protocol (TCP) connections. Therefore, it is intriguing to study the performance of strategies implemented in conventional TCPs, which are not dedicated for video streaming, e.g., whether chunks are efficiently delivered when users perform interactions with the video players. In this paper, we conduct measurement studies on users chunk requesting traces in DASH from a representative video streaming provider, to investigate users behaviors in DASH, and TCP-connection-level traces from CDN servers, to investigate the performance of TCP for DASH. By studying how video chunks are delivered in both the \emph{slow start} and \emph{congestion avoidance} phases, our observations have revealed the performance characteristics of TCP for DASH as follows: (1) Request patterns in DASH have a great impact on the performance of TCP variations including \emph{cubic}; (2) Strategies in conventional TCPs may cause user perceived quality degradation in DASH streaming; (3) Potential improvement to TCP strategies for better delivery in DASH can be further explored.

% \begin{IEEEkeywords}
% DASH; TCP; Congestion Control; Measurement
% \end{IEEEkeywords}

%\textbf{Keywords-DASH; TCP; RTT; bitrate; CWND; congestion control}

\end{abstract}

\section{Introduction}

Recent years have witnessed the increasing popularity of DASH \cite{mpegdash2010}, which allows users with heterogeneous networks and devices to receive video streaming with satisfactory quality-of-experience (QoE). Powered by the infrastructure of Content Delivery Networks (CDN), DASH uses standard HTTP requests for chunk delivery. Since such HTTP requests and responses are based on TCP, it is intriguing to study whether the strategies in conventional TCP variations (e.g., cubic) are effective and efficient for chunk delivery in DASH \cite{tcp_dash}.

\begin{figure}[t]
	\centering
		\includegraphics[width=.8\linewidth]{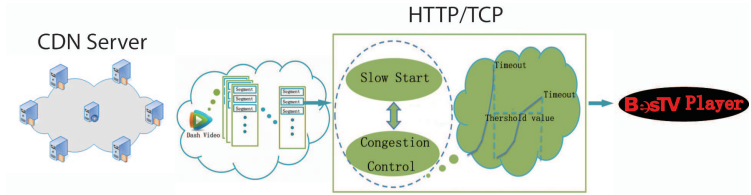}
	\caption{A demonstration of TCP activity in DASH.}
	% \vspace{-0.3cm}
	\label{fig:dashdemo}
\end{figure}

Generally, DASH videos in the CDN servers requested by users will firstly be segmented into different sizes of chunks, and chunks are then delivered over TCP which goes through two phases, \emph{slow start} and \emph{congestion avoidance} \cite{stevens1997tcp}, in a video session. Finally, the chunks are parsed by DASH clients, as illustrated in Fig.~\ref{fig:dashdemo}. To deliver a video chunk, a TCP connection is established between the CDN server and the client. By maintaining the size of the sending window, the server will not over-send packets that can not be received by the client. For quick access to network resources at the beginning, the window size generally grows faster in the slow start phase than in the congestion avoidance phase. In DASH, there are many typical patterns for users to download video chunks, e.g., a user can download chunks intermittently, instead of downloading continuously in large file transmission; and the download intervals change over time, due to the bitrate switch and the dynamic network environment.

% \begin{figure}[t]
% 	\begin{minipage}[t]{0.59\linewidth}
% 		\centering
% 			\includegraphics[width=\linewidth]{ready.eps}
% 			\caption{A demonstration of TCP activity in DASH.}
% 	\label{fig:dashdemo}
% 	\end{minipage}
% 	\hfill
% 	\begin{minipage}[t]{0.39\linewidth}
% 		\centering
% 			\includegraphics[width=\linewidth]{research.eps}
% 			\caption{Framework of our measurement study.}
% 	\label{fig:research}
% 	\end{minipage}
% \end{figure}

On one hand, such chunk request patterns may affect the effectiveness of TCP strategies significantly and on the other hand, one-size-fits-all strategies in conventional TCPs also affect the quality of user experience in DASH. Based on extensive traces on how users request chunks in a large DASH system, and connection-level traces on how CDN servers serve clients, we are able to investigate users behaviors in DASH and measure the performance of TCP for DASH. Our observations reveal not only the mutual effect between TCP strategies and chunk request patterns in DASH, but also the potential improvement that can be conducted to both phases for a better streaming video quality. In particular, our contributions can be summarized as follows.

$\rhd$ \emph{Download behaviors in DASH affect TCP performance.} In our measurement study, we observe that chunk request strategies are designed independently without the awareness of the TCP strategies. (1) Many small flows are generated by today’s DASH players, which request the meta files (e.g., a .m3u8 file) and chunks with very small bitrates frequently, and such small flows affect the overall performance of TCP strategies in both phases; (2) Download patterns are changing over time. Due to the users’ interactions and the dynamic bitrate selection, the TCP performance is far from expected.

$\rhd$ \emph{Possible performance degradation of TCP for chunk delivery in DASH.} Strategies in TCP are not dedicated for DASH streaming either, leading to the user perceived quality degradation. (1) Slow resource allocation. When users perform a sudden player interaction, e.g., seeking, a new TCP connection is established to download the demanded chunks which usually encounters a slow download speed; (2) Bitrate fluctuation. In DASH, bitrate is assigned according to the download speed dynamically. Since the player is not able to predict the download speed of the next TCP connection accurately, the bitrate changes frequently; (3) Flow competition. Since TCP does not guarantee QoS, a DASH connection needs to compete against other flows, leading to quality degradation including bitrate fluctuation and unfair bandwidth sharing at the user side.

$\rhd$ \emph{Insights on improving TCP strategies for DASH.} Based on the observations in the extensive measurement study, we further discuss the possible improvement that we can do to enhance the performance of DASH videos transmission in the two phases of TCP.

	% This paper provides measurements of download speed or delivery time on slow start and congestion avoidance with various congestion window size, RTT, packet loss rate, and different chunk sizes, and geographic locations based on the reasons mentioned above and

% \textbf{Our contributions summary:}
% Our contributions can be summarized as follows: (1) We study the download behaviors when users are watching videos in DASH; (2) We conduct measurement studies to investigate the performance of TCP in delivering the DASH chunks; (3) We study the potential improvement for the chunk delivery strategies.

The rest of the paper is structured as follows. We present the background and our measurement results in Sec.~\ref{sec:measurement}. We present the lessons learnt from the measurement studies in Sec.~\ref{sec:tcpChunk}. We survey related works in Sec.~\ref{sec:relatedwork}. Finally, we conclude the paper with a discussion on the potential improvement for DASH in Sec.~\ref{sec:conclusion}.

\section{Chunk Download Patterns in DASH} \label{sec:measurement}

\begin{figure}[t]
	\centering
		\includegraphics[width=.6\linewidth]{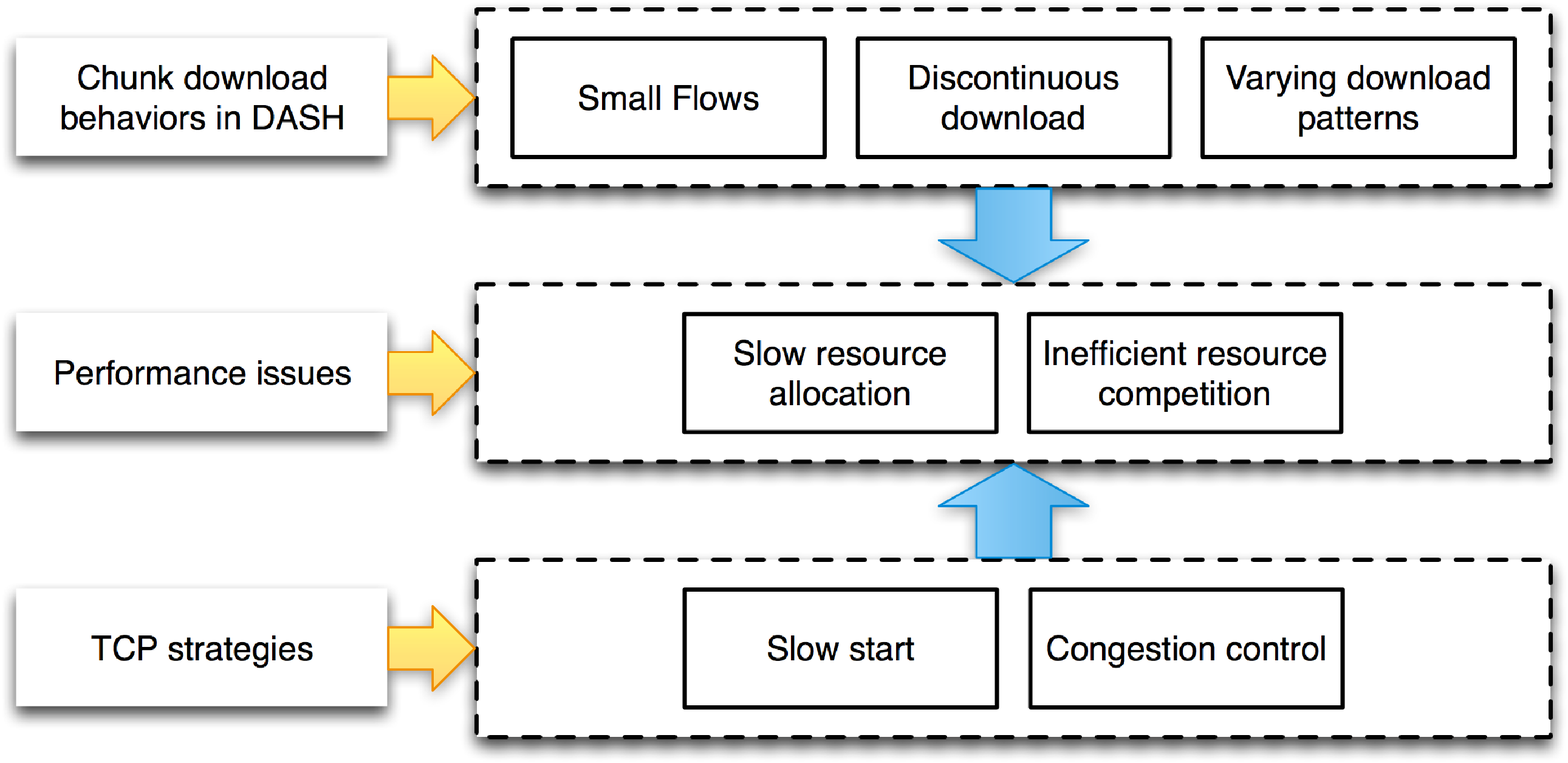}
	\caption{Framework of our measurement study.}
		% \vspace{-0.3cm}
	\label{fig:research}
\end{figure}

Before we present the measurement results, we illustrate the framework of our measurement study in Fig.~\ref{fig:research}. To begin with, we study the chunk download patterns in a representative DASH system, including the small flows, discontinuous download, and varying download patterns. Then we study the strategies in TCP and focus on the slow start and congestion avoidance phases, which have the major impact on the performance of chunk delivery for DASH. Finally, we present that such chunk download patterns over TCP strategies lead to chunk delivery issues, including the slow resource allocation and inefficient resource competition. 

% \begin{figure}[t]
% 	\centering
% 		\includegraphics[width=.6\linewidth]{research.eps}
% 	\caption{Framework of our measurement study.}
% 	\label{fig:research}
% \end{figure}

In this section, we study the chunk download patterns in DASH, which will eventually affect the delivery performance over TCP. We have collected session traces from BesTV \cite{bestv}, one of the largest online video providers in China. It is worth noting that the traces consist of the logs recording how users request the video chunks in DASH.

\subsection{Data Collection}

In order to provide a real and in-depth understanding of relationship between chunk download patterns and delivery performance over TCP, we have collected video session traces from BesTV over $5$ month from Jan. $2013$ to May $2013$. This dataset contains about $1,390$ thousand video sessions and $104$ million items. Each item of the BesTV traces recorded how a DASH video chunk was delivered, including the timestamp when the connection was established, the size of the chunk, the time taken to download the chunk, the device information (e.g., OS type) and the bitrate of the chunk, indicated by S1 (avg.~$700$kbps), S2 (avg.~$1300$kbps), S3 (avg.~$2300$kbps), and S4 (avg.~$4000$kbps).

We also have collected real world TCP-connection-level traces of DASH video delivery from Tencent \cite{tencentvideo} from Jun. $2013$ to Aug. $2013$. These traces are collected from two servers, which are dedicated for video delivery, of Tencent in Shenzhen, China. Note that the two servers are deployed by Tencent for DASH video delivery and under our control. We adopt different strategies to adjust congestion window (CWND) sizes for the servers. One server is under default TCP and another adopts an intelligent and adaptive algorithm proposed in \cite{wang2011tcp} to adjust CWND size according to network status and chunk size.

\begin{figure}[t]
	\begin{minipage}[t]{0.32\linewidth}
		\centering
			\includegraphics[width=\linewidth]{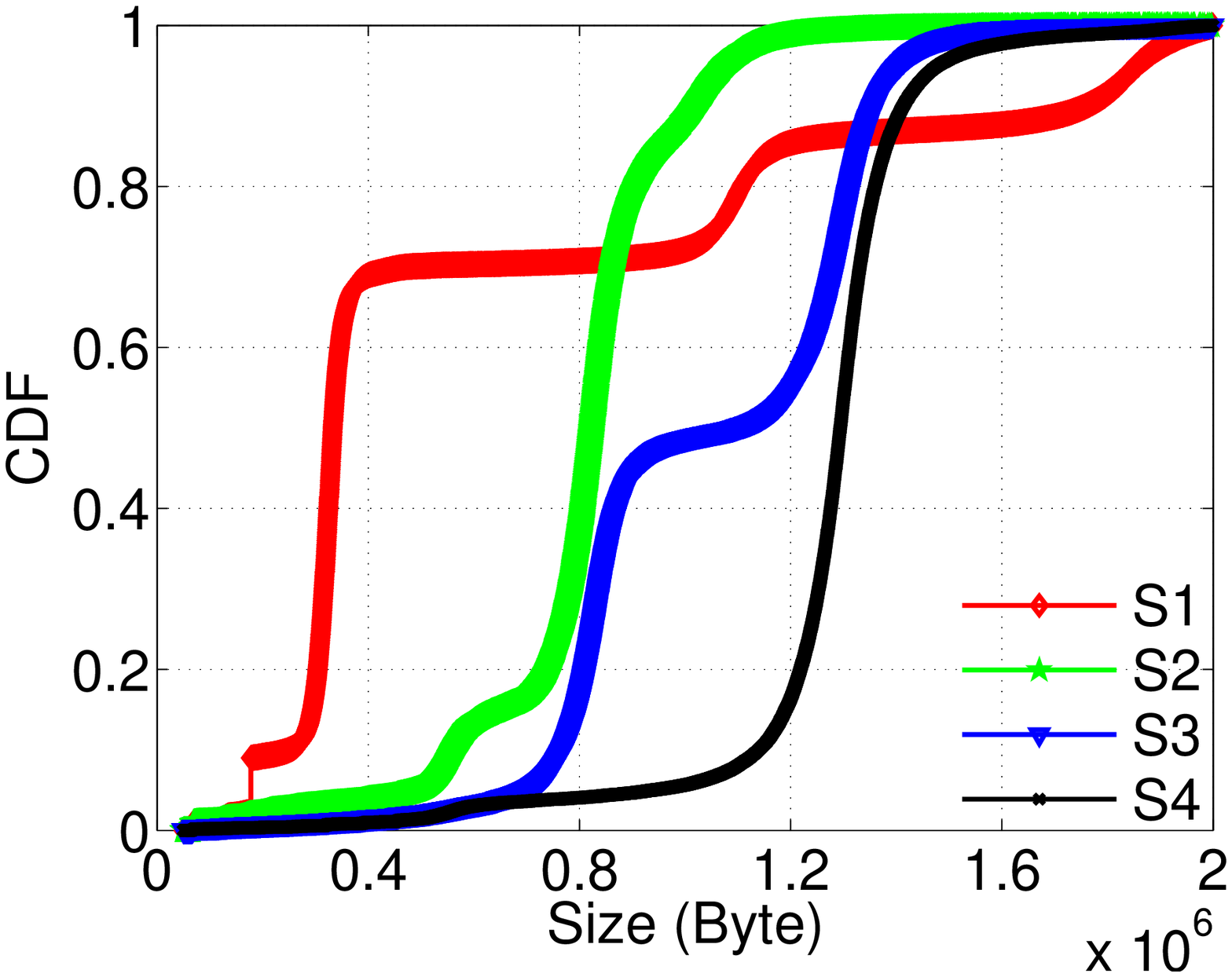}
			\caption{CDF of chunk files size.}
			\label{fig:side:difbitratesize}
	\end{minipage}
	\hfill
	\begin{minipage}[t]{0.32\linewidth}
		\centering
			\includegraphics[width=\linewidth]{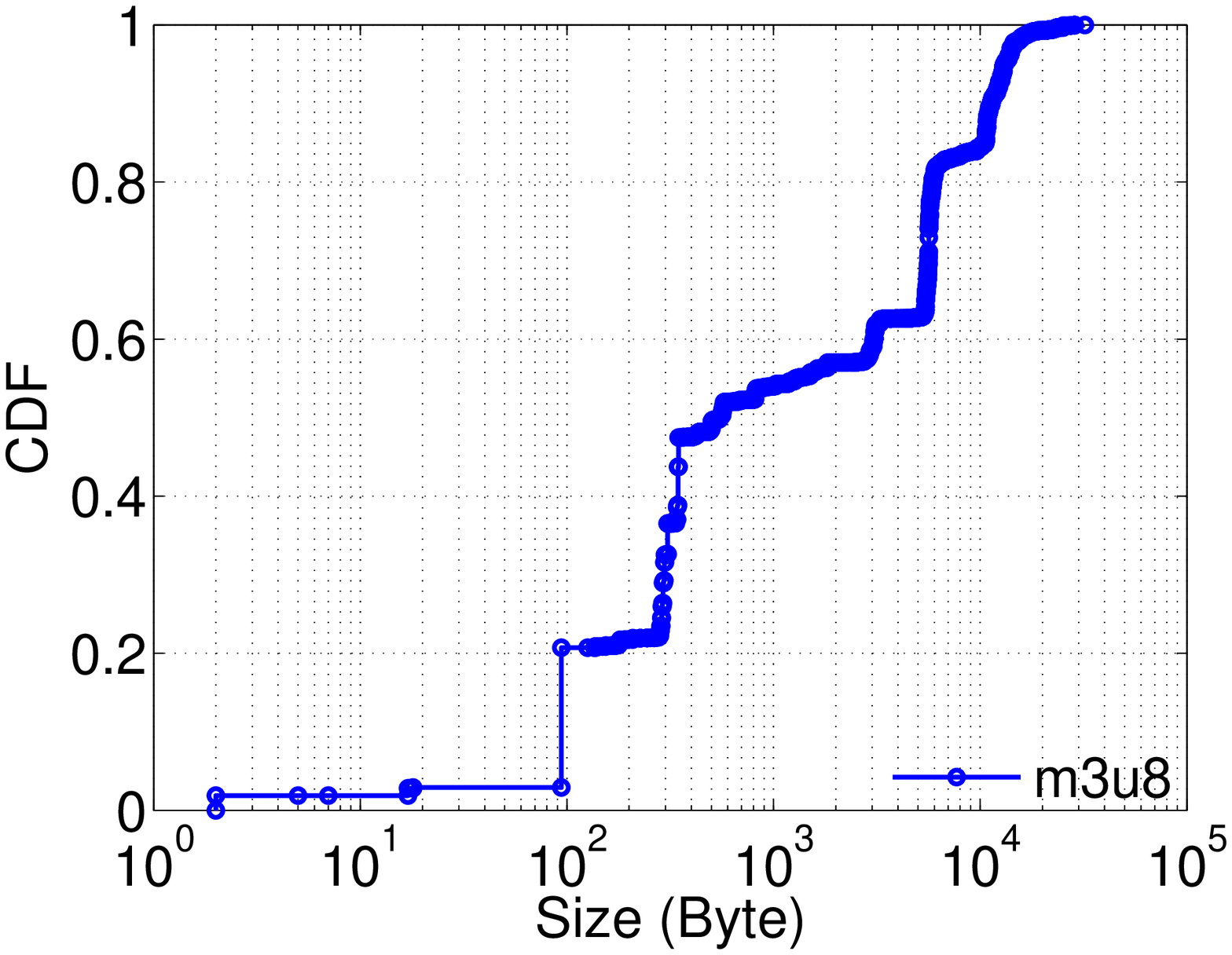}
			\caption{CDF of .m3u8 files size.}
			\label{fig:side:m3u8Size}
	\end{minipage}
	\hfill
	\begin{minipage}[t]{0.32\linewidth}
		\centering
			\includegraphics[width=\linewidth]{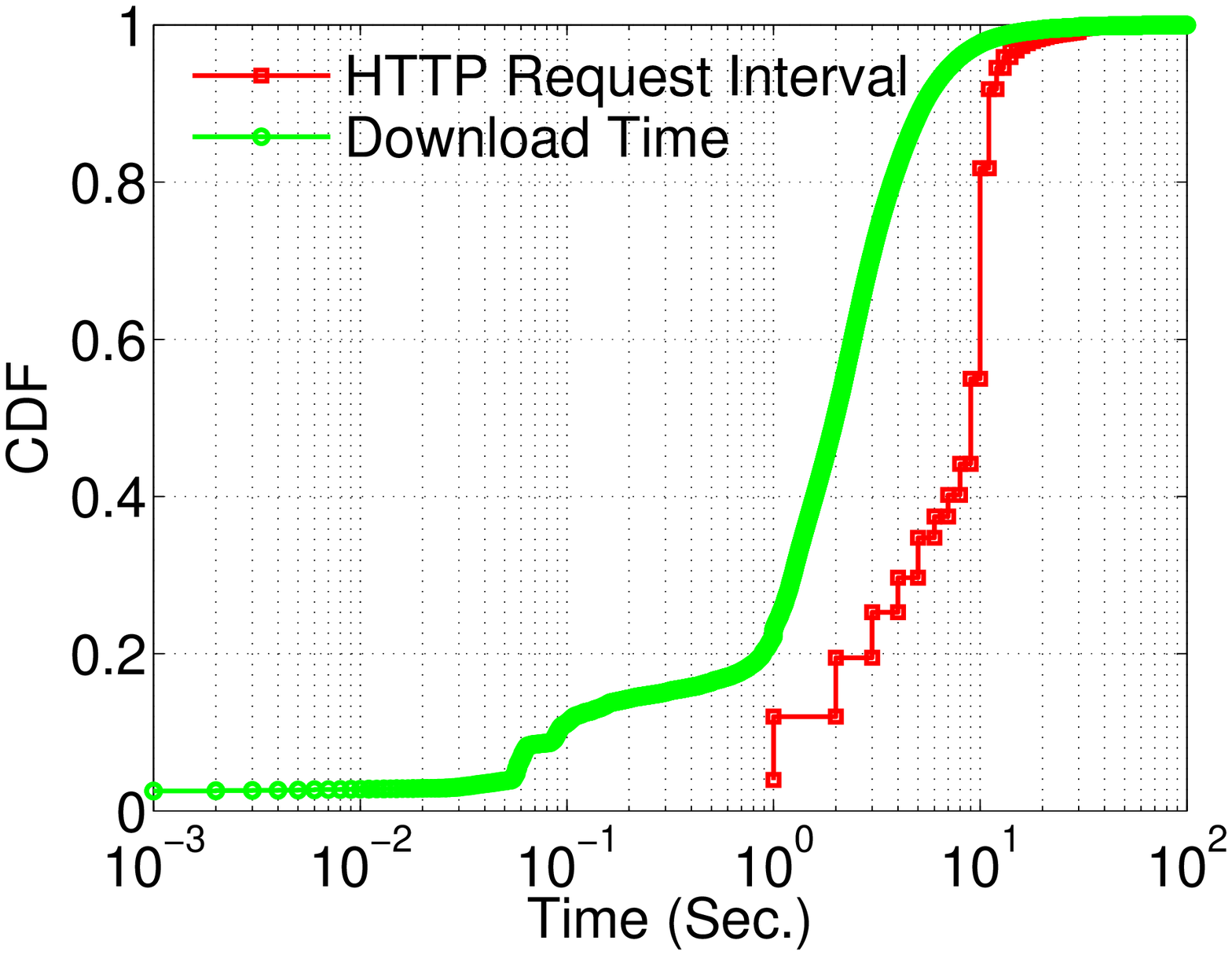}
			\caption{CDF of HTTP request interval and download time.}
		\label{fig:IntervalHttpRequestandDownload}
	\end{minipage}
\end{figure}
%
%\begin{figure}[t]
%	\centering
%		\includegraphics[width=0.8\linewidth]{DifBitrateSize.eps}
%	\caption{CDF of chunk files size.}
%	\label{fig:difbitratesize}
%\end{figure}

%\begin{figure}[t]
%	\centering
%		\includegraphics[width=0.8\linewidth]{m3u8Size.eps}
%	\caption{CDF of m3u8 files size.}
%	\label{fig:m3u8Size}
%\end{figure}%

\subsection{Small Flows}

We study the size of flows in chunk delivery in DASH sessions. Chunk size is statistically related to the bitrate because playback time is almost the same according to the traces. As illustrated in Fig.~\ref{fig:side:difbitratesize}, each curve is the cumulative distribution function (CDF) of the flow size for a particular DASH bitrate version. We observe that the size of all the chunks delivered is smaller than $2$MB. In particular, for the bitrate version S1, the size of flows is smaller than $400$KB mostly. Besides, DASH meta files (i.e., .m3u8 file in BesTV) are frequently requested by DASH players, and the size of the .m3u8 files is much smaller than video chunks. As illustrated in Fig.~\ref{fig:side:m3u8Size}, over $53\%$ (resp. $99.5\%$) of .m3u8 files have a size smaller than $1$KB (resp. $30$KB). In summary, TCP has to handle small flows for DASH.

Those small flows can affect the performance of TCP strategies greatly in the slow start and congestion avoidance phases, since the network resource has to be allocated in a very instantaneous manner for the delivery of such small flows. 

% \begin{figure}[t]
% 	\centering
% 		\includegraphics[width=0.8\linewidth]{IntervalHttpRequestandDownload.eps}
% 	\caption{CDF of HTTP request interval and download time.}
% 	\label{fig:IntervalHttpRequestandDownload}
% \end{figure}
%
%
% \begin{figure}[t]
% 	\centering
% 		\includegraphics[width=0.8\linewidth]{DifBitrateInterval.eps}
% 	\caption{CDF of download intervals of chunks and metafiles.}
% 	\label{fig:DifBitrateInterval}
% \end{figure}

\subsection{Discontinuous Chunk Download}

\begin{figure}[t]	
	\begin{minipage}[t]{0.32\linewidth}
		\centering
			\includegraphics[width=\linewidth]{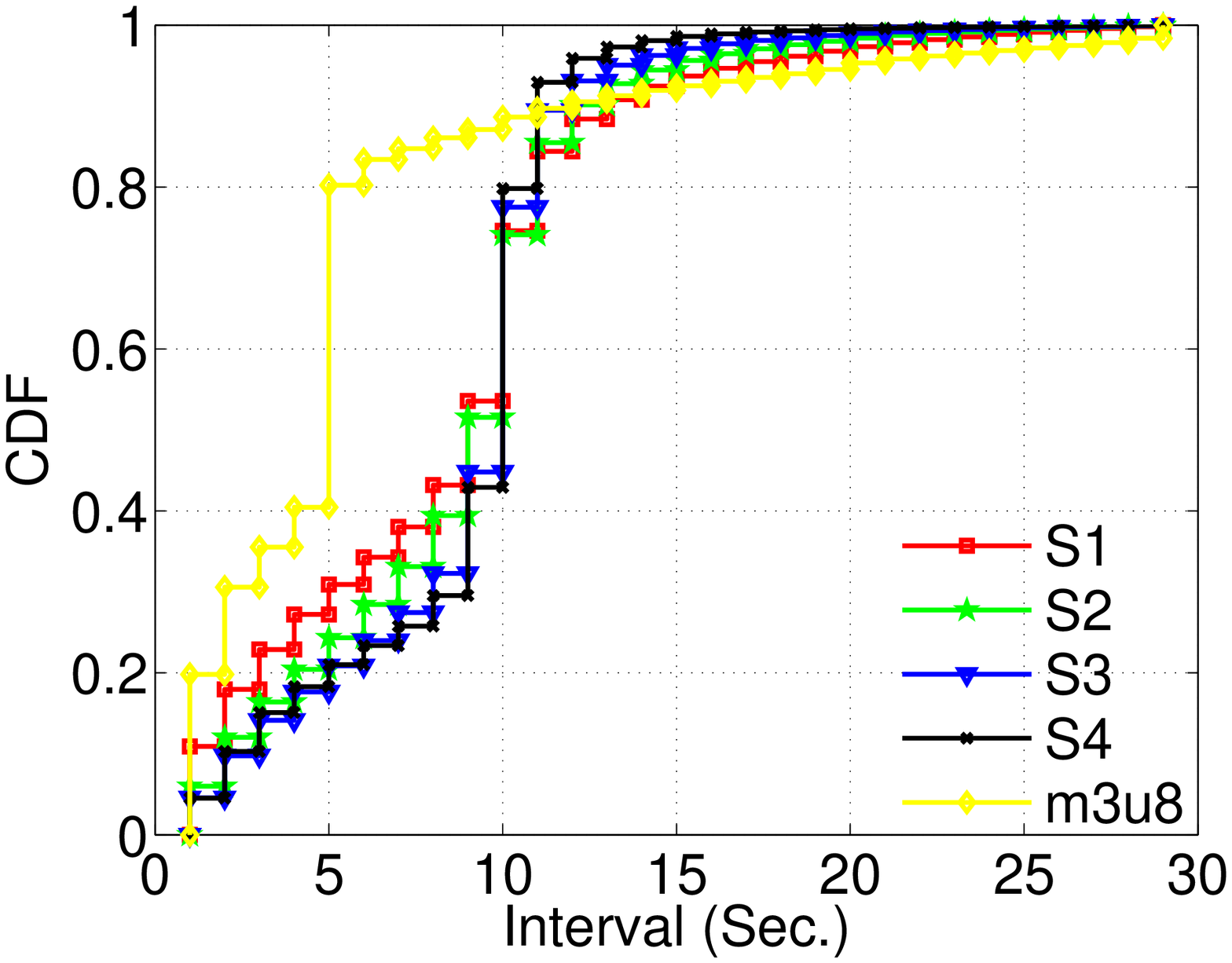}
			\caption{CDF of download intervals of chunks and metafiles.}
			\label{fig:DifBitrateInterval}
	\end{minipage}
	\hfill
	\begin{minipage}[t]{0.32\linewidth}
		\centering
			\includegraphics[width=\linewidth]{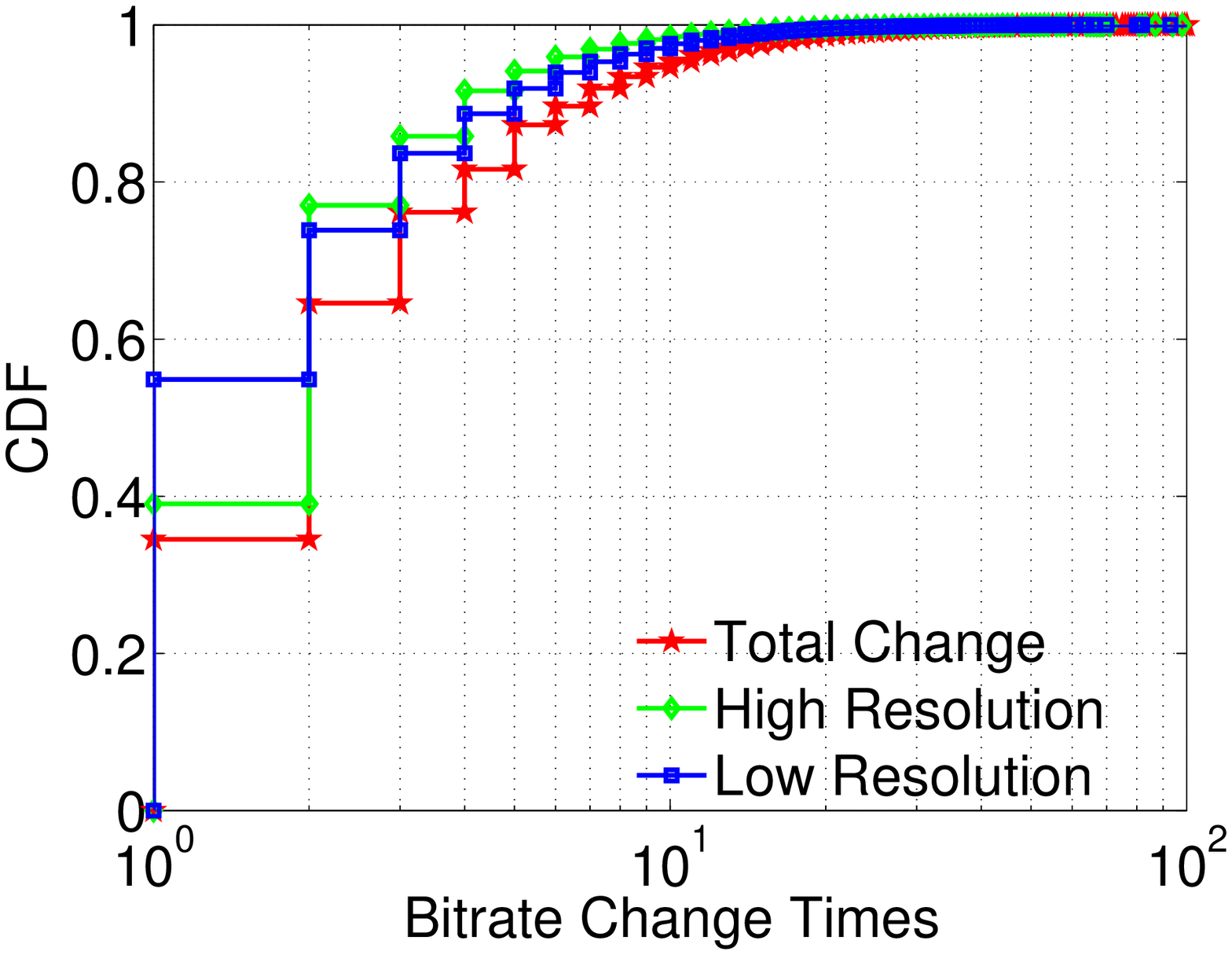}
			\caption{CDF of bitrate change times.}
			\label{fig:BitrateConversionTimes}
	\end{minipage}
	\hfill
	\begin{minipage}[t]{0.32\linewidth}
		\centering
			\includegraphics[width=\linewidth]{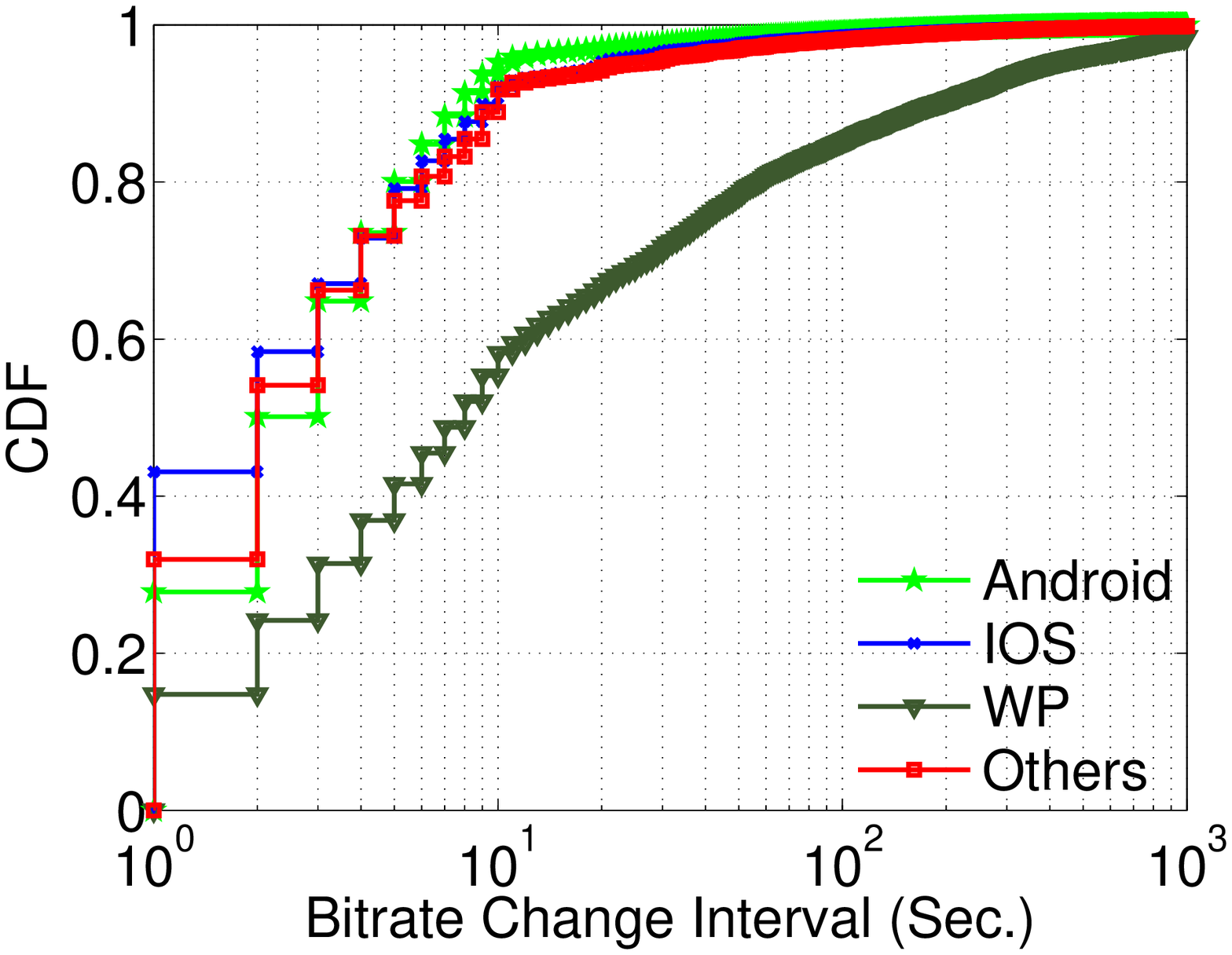}
			\caption{CDF of bitrate change interval.}
			\label{fig:BitrateConversionInterval}
	\end{minipage}
\end{figure}

In our study, we also observe that DASH video chunks are downloaded discontinuously. As illustrated in Fig.~\ref{fig:IntervalHttpRequestandDownload}, the curves denote the CDF of chunk request interval (i.e., the average time elapse between two consecutive chunk requests), and the CDF of the download time (i.e., the time used to download a chunk) respectively. (1) We observe that over $98\%$ (resp. $20\%$) of the video chunks are downloaded within $10$ seconds (resp. $1$ second). (2) We observe that nearly $50\%$ of the chunk request interval is around $10$ seconds and the download time is much smaller than the chunk request interval, indicating that the downloads take place discontinuously. (3) Furthermore, for more detailed analysis, we decompose the chunk request intervals into different bitrate versions, as illustrated in Fig.~\ref{fig:DifBitrateInterval}. We observe that such discontinuous downloads exist in both the meta files (.m3u8) and chunks with different bitrates, and chunks with higher bitrate tend to have a relative concentration interval, which is consistent with users' watching experience. Users expect to obtain high speed and stable network when they watch high quality DASH videos, and that network in turn ensures a higher proportion of request interval concentrates on a small scale (around $10$ second) as shown in Fig.~\ref{fig:DifBitrateInterval}.

% So we can conclude that servers of DASH videos offer the best service of quality under the current network conditions. We also obtain that when users download the same video, the higher of HTTP request frequency, the greater the average download speed under the same network condition (The result is not shown in this paper). 

\subsection{Changing Request Patterns over Time}

Due to both the users' interactions and the dynamic bitrate selection, DASH video chunks' download patterns are changing over time, which also makes the TCP performance far from expected. Before diving into the details, we define several bitrate ``changes'' as follows. (1) \emph{High resolution change} indicates that the client requests to download a chunk with a higher bitrate than the previous chunk request. (2) \emph{Low resolution change} indicates that the client requests to download a chunk with a lower bitrate than the previous chunk request.

First, we study the changes in bitrates when users download chunks in DASH sessions. As illustrated in Fig.~\ref{fig:BitrateConversionTimes}, each curve is the CDF of the number of bitrate changes in a session. We observe that over $80\%$ of the sessions have a change number smaller than $4$, which is largely caused by short video playback time, while there is still a certain fraction of sessions with frequent bitrate changes.

Second, we investigate the intervals between consecutive bitrate changes. As illustrated in Fig.~\ref{fig:BitrateConversionInterval}, the curves represent the CDF of intervals between consecutive bitrate changes in DASH sessions for different types of operating system (OS) devices. We have the following observations: (1) It is consistent with the previous result, that requesting patterns are changing over time; (2) For different OS devices, it seems that the distribution of bitrate change interval is different. The reason may be that DASH players in different OS devices are designed with different strategies to change the bitrate selection, but this is not our focus. We do observe that nearly $90\%$ of conversion interval of three different OSes (Android, iOS, and others) and $60\%$ of WP (Windows Phone) is within 10 seconds, which further indicates that bitrate changes happen in a short time and frequently.

%In the process of calculating the bitrate conversions, we make such an assumption based on the HTTP requests' interval previously as illustrated in Fig.~\ref{fig:DifBitrateInterval}: when the adjacent two HTTP requests' interval is longer than 30 Seconds, we believe the current session ends. 

% \begin{figure}[t]
% 	\centering
% 		\includegraphics[width=0.8\linewidth]{UPAndDownResolution.eps}
% 	\caption{CDF of bitrate change times.}
% 	\label{fig:BitrateConversionTimes}
% \end{figure}
%
% \begin{figure}[t]
% 	\centering
% 		\includegraphics[width=0.8\linewidth]{DifDEVBitrateInterval.eps}
% 	\caption{CDF of bitrate change interval.}
% 	\label{fig:BitrateConversionInterval}
% \end{figure}

\section{Performance of Chunk Delivery over TCP} \label{sec:tcpChunk}

In this section, we study the performance of TCP strategies, when chunks are delivered according to the request patterns studied in the previous section. As illustrated in Fig.~\ref{fig:dashdemo}, when a chunk is delivered over TCP, the initial CWND size and congestion avoidance affect the transmission of DASH videos. In particular, We will study their performance for delivering chunks.

\subsection{Slow Resource Allocation in Delivering Small Files}

There are many cases in a DASH session that will cause a competition for network resource in the slow start phase, e.g., (1) When the client downloads meta files (e.g., .m3u8 files) or small chunks, the initial congestion window size determines the download performance. (2) When users perform a sudden player interaction, a new TCP connection is established to download the demanded chunks which usually encounters a slow download speed. Since CWND determines the performance of TCP in the slow start phase, we study the impact of CWND on the chunk/meta file delivery in DASH.

% \begin{figure}
% \begin{minipage}[t]{0.49\linewidth}
% \centering
% \includegraphics[width=\linewidth]{DeliveryTimeVersusCWND.eps}
% \caption{Delivery time versus initial CWND.}
% \label{fig:side:delivery-time-vs-cwnd}
% \end{minipage}
% \hspace{0.005in}
% \begin{minipage}[t]{0.49\linewidth}
% \centering
% \includegraphics[width=\linewidth]{SizeOfInitialCWND-ms.eps}
% \caption{Download speed versus initial CWND.}
% \label{fig:side:download-speed-vs-cwnd}
% \end{minipage}
% \end{figure}

% \begin{figure}[t]
% 	\centering
% 		\includegraphics[width=0.8\linewidth]{DeliveryTimeVersusCWND.eps}
% 	\caption{Delivery time versus initial CWND.}
% 	\label{fig:side:delivery-time-vs-cwnd}
% \end{figure}

%\begin{figure}\centering
	
%	\subfigure{
%		\label{fig:side:delivery-time-vs-cwnd}
%		\includegraphics[width=0.5\linewidth]{DeliveryTimeVersusCWND.eps}
%		%\caption{Delivery time versus initial CWND.}
%	}		
	
%	\subfigure{
%		\label{fig:side:download-speed-vs-cwnd}
%		\includegraphics[width=0.5\linewidth]{SizeOfInitialCWND-ms.eps}
		%\caption{Download speed versus initial CWND.}
%	}
%	\caption{Download speed versus initial CWND.}

%\end{figure}

\begin{figure}[t]
	\begin{minipage}[t]{0.32\linewidth}
		\centering
			\includegraphics[width=\linewidth]{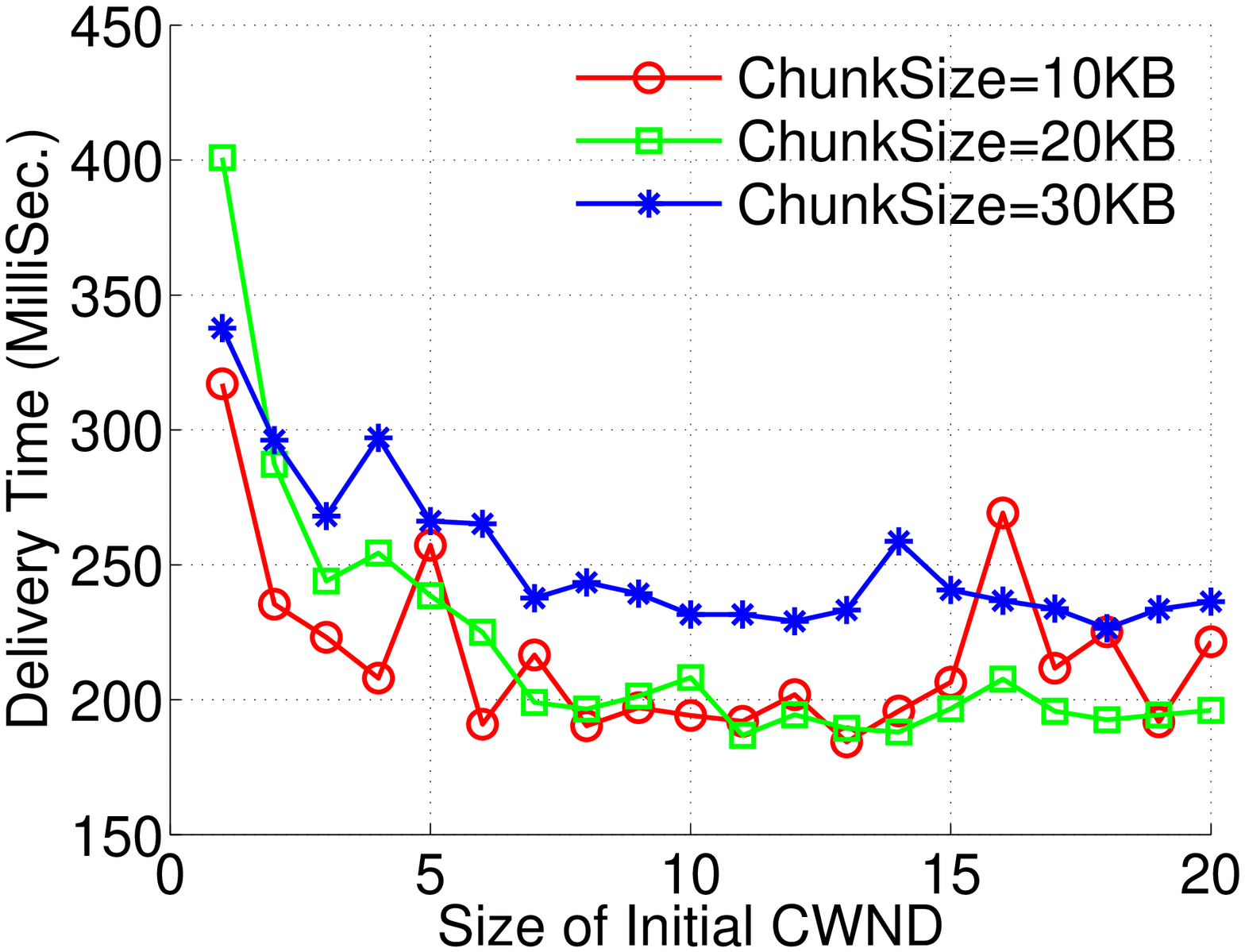}
		\caption{Delivery time versus initial CWND.}
		\label{fig:side:delivery-time-vs-cwnd}
	\end{minipage}
	\hfill
	\begin{minipage}[t]{0.32\linewidth}
		\centering
			\includegraphics[width=\linewidth]{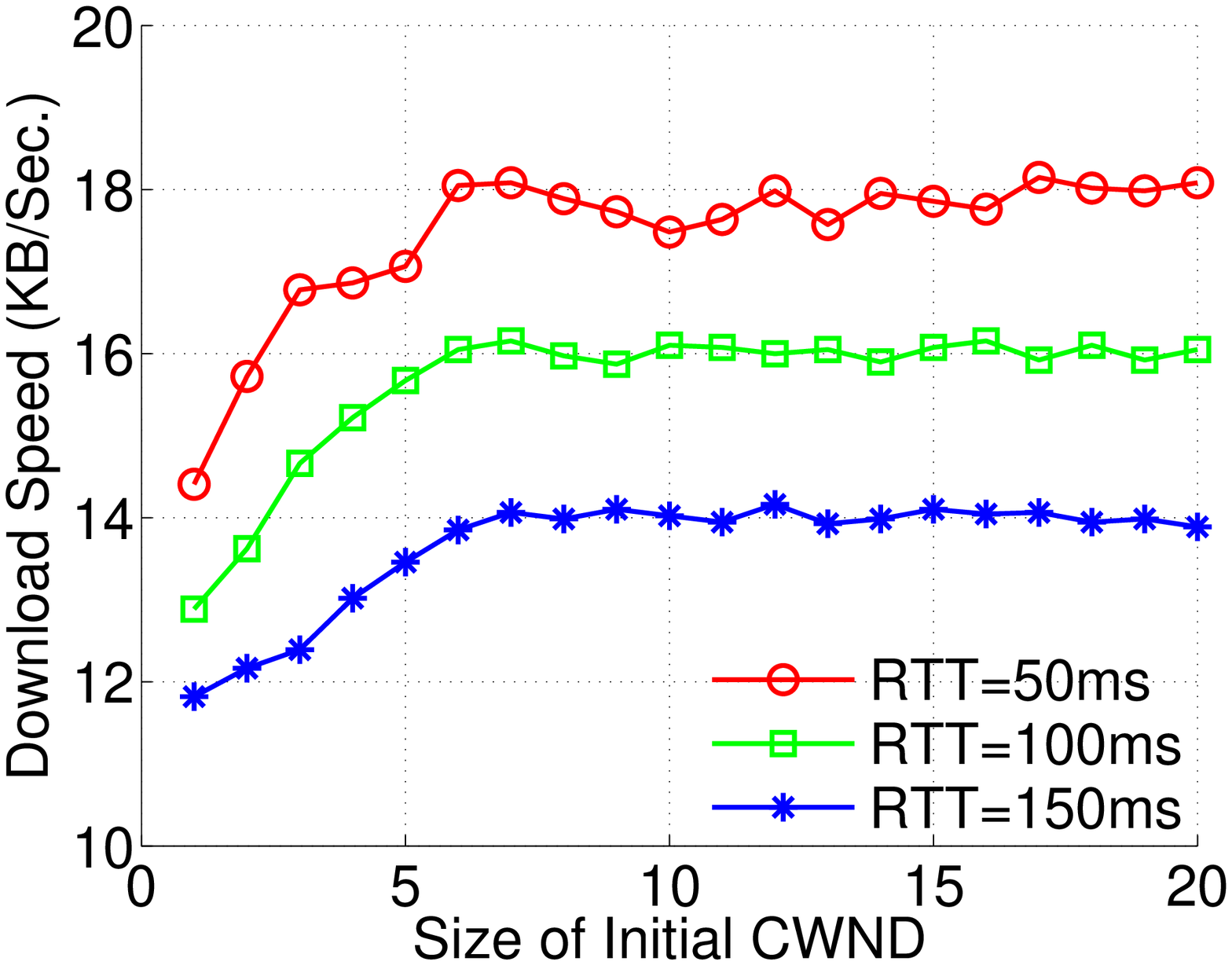}
			\caption{Download speed versus initial CWND.}
	\label{fig:side:download-speed-vs-cwnd}
	\end{minipage}
	\hfill
	\begin{minipage}[t]{0.32\linewidth}
		\centering
			\includegraphics[width=\linewidth]{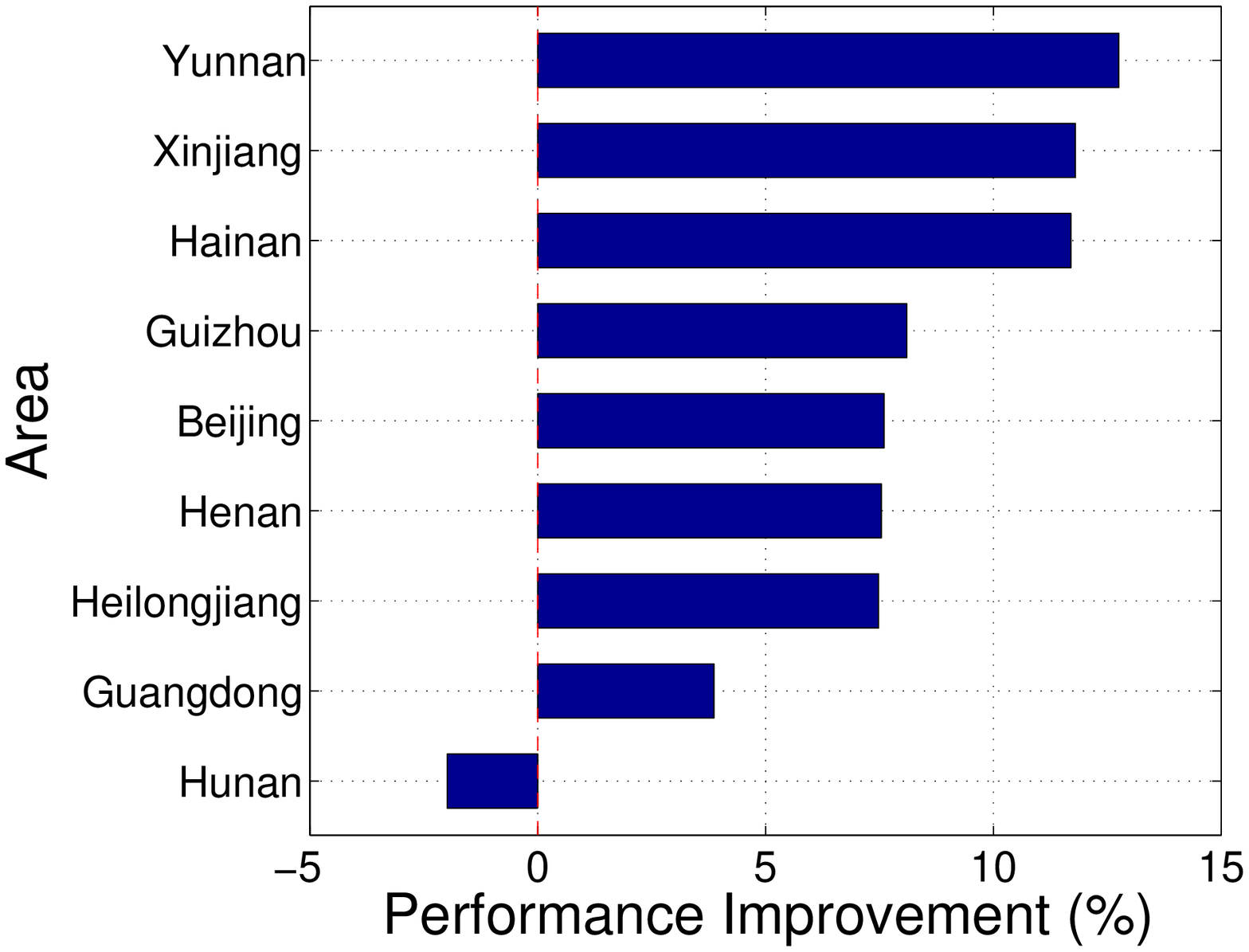}
			\caption{Performance improvement of different areas.}
			\label{fig:performanceimprovement}
	\end{minipage}
\end{figure}

\textbf{Results in the wild.} Based on the TCP traces, we have got the average delivery time for video chunks in DASH. Fig.~\ref{fig:side:delivery-time-vs-cwnd} compares the average delivery time of a chunk with different sizes, under different initial CWND sizes in the real-world network environment. In this figure, each sample represents the average delivery time versus the initial CWND size. We have made the following observations: (1) There is a general trend that a larger initial CWND leads to less delivery time, i.e., the average delivery time has been reduced by $50\%$ when the initial CWND grows from $1$ to $10$. (2) The initial CWND tends to have a larger impact on smaller chunks, e.g., for the chunk size of $10$KB, the delivery time is reduced by about $50\%$, while for the chunk size of $30$KB, the same initial CWND increment reduces the delivery time by only about $25\%$. (3) We observe that when the initial CWND is large enough, increasing it continuously will not reduce the delivery time any more. The reason is that the DASH chunk transmission goes to the other phase, the congestion avoidance phase. And we conduct another experiment that we download two group files from Beijing to Shenzhen respectively. One group is one thousand chunks with $1$MB, and another is two thousands chunks with $500$KB. Transmission of those files must go through congestion avoidance phase and we find that average download speed of files with $500$KB is $23.08$\% improved compared to that of files with $1$MB. The results above can confirm the conclusion that request patterns of DASH players have a significant impact on the performance of the TCP congestion controls.

%It is quite natural that we can also take a emulation on the different chunk size and initial CWND size with the same RTT, as illustrated in Figure 4. We can easily get that increasing the size of initial CWND in a certain range can reduce the delivery time distinctly, i.e., the average delivery time has been reduced by more than 15.8\% when initial CWND raised from 1 to 7. When size of initial CWND beyond the scope of 7, the average delivery time is almost the same. Each curve is identically distributed regardless of the expectation value of delivery time.

% \begin{figure}[t]
% 	\centering
% 		\includegraphics[width=0.8\linewidth]{SizeOfInitialCWND-ms.eps}
% 	\caption{Download speed versus initial CWND.}
% 	\label{fig:side:download-speed-vs-cwnd}
% \end{figure}

\textbf{Results in the controlled experiments.} In order to eliminate the impact of changing round-trip time (RTT) on the download speed, we also run experiments in a controlled platform (i.e., PlanetLab) to demonstrate the correlation between download speed and the initial CWND. The results are illustrated in Fig.~\ref{fig:side:download-speed-vs-cwnd}. All samples are with the same chunk size and RTT, under different initial CWND sizes. We observe similar results, i.e., the download speed is increasing along with the initial CWND size. In particular, an improvement of download speed by $25.6\%$ is observed, when the initial window size improves by $3\sim4$, and RTT is $50$ms. However, when the initial CWND size is larger than $6$, the download speed tends to level off.

% \begin{figure}[t]
% 	\centering
% 		\includegraphics[width=0.8\linewidth]{performance.eps}
% 	\caption{Performance improvement of different areas.}
% 	\label{fig:performanceimprovement}
% \end{figure}

\textbf{Performance improvement.} Based on the results in the wild and controlled experiment, we conduct real measurements by configuring two servers with the same setting except the CWND size strategies adopted according to the results as mentioned in Sec.~\ref{sec:measurement}. We assume that the connections between clients and servers are independent and random, and TCP performance and delivery speed of DASH video are positive correlated. Then we can study the download speed simply to investigate the TCP performance for DASH. Let us define the average download speed of the server with default TCP and the server with adaptive CWND size as $AverageServer1  = \frac{\sum_{n=1}^{N} SpeedServer1(n)}{N}$ and $AverageServer2 = \frac{\sum_{m=1}^{M} SpeedServer2(m)}{M}$,
% $$
% AverageServer1  = \frac{\sum_{n=1}^{N} SpeedServer1(n)}{N}
% $$
%
% $$
% AverageServer2 = \frac{\sum_{m=1}^{M} SpeedServer2(m)}{M}
% $$
where $\mathit{SpeedServer1}$ (resp. $\mathit{SpeedServer2}$) is the download speed of server $1$ (resp. server $2$) for each connection, and $\mathit{N}$ and $\mathit{M}$ are the corresponding number of samples. Then the performance improvement is defined as $Improvement = \frac{AverageServer2 - AverageServer1}{AverageServer1}$.
% $$
% Improvement = \frac{AverageServer2 - AverageServer1}{AverageServer1}
% $$.

%\begin{gather}
%\doublebox {$Improvement = \dfrac{AverageServer2 - AverageServer1}{AverageServer1}$}
%\end{gather}
Fig.~\ref{fig:performanceimprovement} shows that the performance improves significantly as expected in different areas except Hunan, and $12.75$\% is achieved typically in Yunnan. We then investigate the reason why negative results happened in Hunan and find that the samples is only $2$\% of the total number in nine provinces, which is too small to have statistical significance.

\subsection{Resource Competition in Chunk Delivery}

\begin{figure}[t]
	\begin{minipage}[t]{0.48\linewidth}
		\centering
			\includegraphics[width=.9\linewidth]{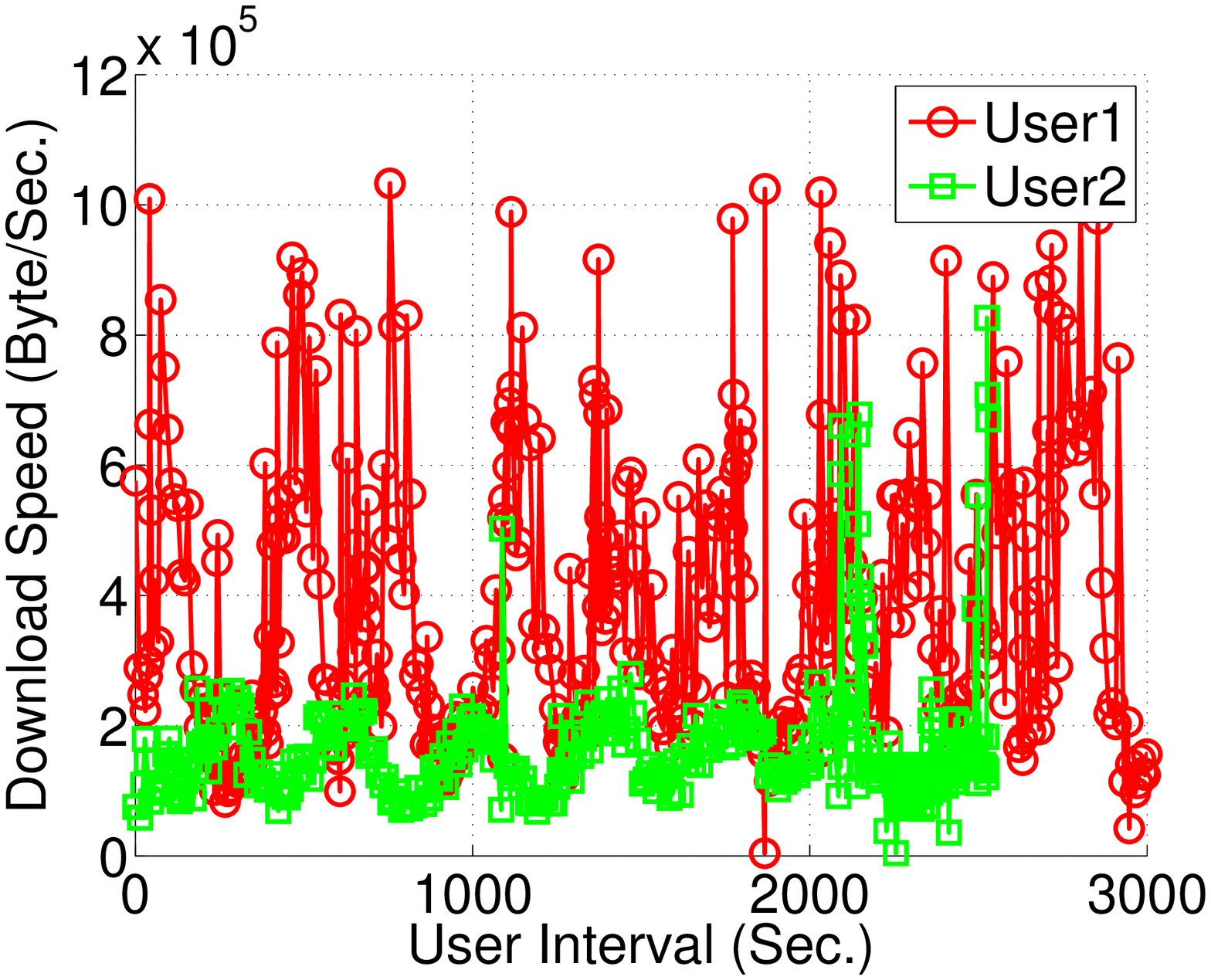}
		\caption{Download speed over time.}
	\label{fig:speedlog}
	\end{minipage}
	\hfill
	\begin{minipage}[t]{0.48\linewidth}
		\centering
			\includegraphics[width=.9\linewidth]{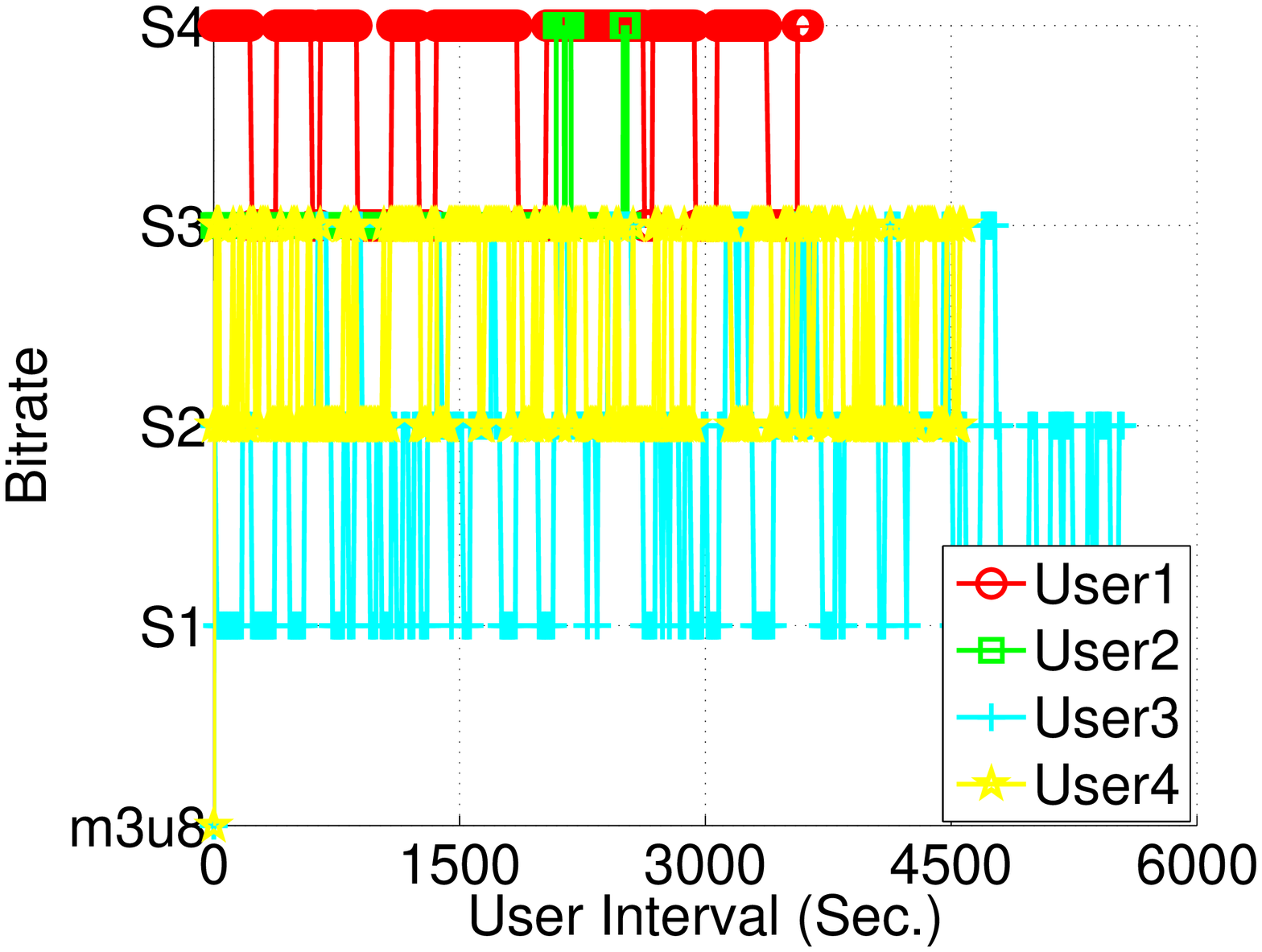}
		\caption{Bitrate fluctuation over time.}
	\label{fig:userlog}
	\end{minipage}
\end{figure}

First, we study the resource competition in TCP when users download via a bottleneck. Two users competed for a bottleneck with each other are selected from our traces. As illustrated in Fig.~\ref{fig:speedlog},  we set the start time to $0$ and the curves represent the download speed of two users. We have made the following observations: (1) Download speed jitter is very intense, e.g., the download speed of user1 changes in a range of $50$KBps to $1$MBps. (2) There is a obvious negative correlation between two users' download speed. We observe that the download speed of user1 increases rapidly after user2 experiences the first speed peak. The results indicate that TCP is not designed with QoS guarantee. Let users compete bandwidth resource at the bottlenecks may have a great impact on the streaming quality in DASH. TCP should adjust the bandwidth allocation to improve QoE for users based on the dynamic characteristics of the video and the current bitrate, although bottleneck link bandwidth competition is a general problem for all types of traffic.

Since TCP does not guarantee QoS, a DASH connection needs to compete against other flows, resulting in a changing quality of streaming at the users. In our traces, we have collected multiple users with the same IP address (e.g., in a NAT) requesting chunks during the same period. As illustrated in Fig.~\ref{fig:userlog}, such users are competing resource at the bottleneck close to them, e.g., the local downlink capacity. We observe that user3 and user4 are intensively competing for bandwidth, resulting in frequent bitrate changes, e.g., user3 encounters frequent changes between S1 and S2. Our results indicate that as TCP strategies are not streaming aware fundamentally, especially for streaming in DASH with adaptive bitrates, it is hard for users to take full advantage of the dynamic bitrate adaptation.

% \begin{figure}[t]
% 	\centering
% 		\includegraphics[width=0.8\linewidth]{speedlog.eps}
% 	\caption{Download speed over time.}
% 	\label{fig:speedlog}
% \end{figure}
%
% \begin{figure}[t]
% 	\centering
% 		\includegraphics[width=0.8\linewidth]{userlog.eps}
% 	\caption{Bitrate fluctuation over time.}
% 	\label{fig:userlog}
% \end{figure}

\section{Related Work} \label{sec:relatedwork}

\subsection{DASH Streaming and its Chunk Delivery}

Sodagar et al.~\cite{sodagar2011mpeg} defined the MPEG-DASH standard for multimedia streaming over internet and further defined five specific profiles, each of which addressed a different class of applications, and a set of constraints, limiting the MPD and segment formats to a subset of the entire specification. Apple~\cite{appledash2010} presented an overview for HTTP Live Streaming, and introduced the architecture of HTTP streaming and how to use the HTTP Live Streaming in details. Adobe~\cite{adobedash2011} introduced Adobe HTTP Dynamic Streaming and even shared a user guide to tell us how to use the DASH system. Joseph et al.~\cite{joseph2014nova} presented a simple asymptotically optimal online algorithm, NOVA, to optimize video delivery for a network supporting video clients streaming stored video and maximize the QoE of video clients. Akhshabi et al.~\cite{akhshabi2012happens} studied the performance problems when two adaptive streaming players shared the same network bottleneck and competed for the available bandwidth. 
% \cite{el2013quality} proposed a proactive QoE-based approach which could rewrite the client HTTP requests at a proxy over next generation wireless networks, such as 4G. Different from conventional DASH implementing rate adaptation logics either locally in the user equipment or in the DASH server, \cite{pu2012video} developed a wireless DASH (WiDASH) proxy located at the edge network to enhance users' QoE and claimed that the WiDASH proxy could perform global optimization over multiple concurrent DASH flows. \cite{de2014qoe} presented a QoE model for video delivered over a radio network and evaluated the impact of the radio parameters on the video quality perceived by end users. 
However, these works either did not consider the TCP performance for DASH or did not design strategies based on the unique characteristics of DASH, e.g., small flows, discontinuous download, and varying download patterns. % \cite{park2015rate} found that HTTP Adaptive Streaming flow can not fairly share the bottleneck with TCP flows due to the OFF period and TCP dynamics, which can highlight the significance of our measurement study.

%However, these works did not consider the impact of the transport protocol, i.e., whether the TCP was efficient for the DASH chunk delivery. \cite{park2015rate} found that HTTP Adaptive Streaming flow can not fairly share the bottleneck with TCP flows due to the OFF period and TCP dynamics, which can highlight the significance of our measurement study.

\subsection{TCP for Video Chunk Delivery}

% \cite{bhat2014performance} studied how to provide low end-to-end delay while streaming video with TCP as the transport layer and proposed a random linear coding-based set of enhancements to TCP’s flow control called Variable Bucket Size Network Coding, which offered almost a $100$\% improvement in video goodput over TCP New Reno. \cite{hossfeld2014qoe} investigated whether TCP fit Video-on-Demand delivery from the end user’s perspective, when the video was transmitted over a bottleneck link, and found that the displayed content was not disturbed but playback suffered from stalling due to re-buffering. 

% \cite{hacker2004improving} proposed an approach to improve the throughput effectively on an uncongested network and maintain fairness using parallel TCP when the network was congested. \cite{esteban2012interactions} investigated the interplay between HTTP adaptive streaming and TCP, and the impact of network delay on achievable throughput. They divided the transmission into 3 phases, initial burst, acknowledgement (ACK) clocking, and trailing ACK, and claimed that packet losses are the most damaging during the trailing ACK phase and are least disruptive during the ACK clocking phase, which is the basis of our measurement in CWND size and packet loss rate. 

Hacker et al.~\cite{hacker2004improving} proposed an approach to improve the throughput effectively on an uncongested network and maintain fairness using parallel TCP when the network was congested. Esteban et al.~\cite{esteban2012interactions} investigated the interplay between HTTP adaptive streaming and TCP, and the impact of network delay on achievable throughput. 
% They divided the transmission into 3 phases, initial burst, acknowledgement (ACK) clocking, and trailing ACK, and claimed that packet losses are the most damaging during the trailing ACK phase and are least disruptive during the ACK clocking phase, which is the basis of our measurement in CWND size and packet loss rate. 
% \cite{havey2012receiver} presented a novel client driven application layer rate adaptation mechanism to achieve throughput increase compared to a standard rate adaptive HTTP/TCP video stream operating over wireless network.
% \cite{allman1998increasing}, \cite{allman1999tcp} demonstrated the importance of initial window size by simulation.
Allman et al.~\cite{allman1999tcp} studied the advantages and disadvantages if we raised the initial window size and how TCP should begin transmission after a relatively long idle period. However, none of them took the size of chunk to be delivered and network status into consideration. On the other hand, congestion control algorithm has been explored in \cite{wang2011tcp}, \cite{fu2003tcp}, \cite{ha2008cubic}, \cite{wei2006fast}, and \cite{floyd2004newreno}, and a variety of algorithms have been proposed to improve the performance of congestion avoidance. Wang et al.~\cite{wang2013cubic} proposed a new congestion control algorithm, which was more efficient for data delivery in the case of long distance and wireless network than other algorithms such as TCP CUBIC \cite{ha2008cubic} and Veno \cite{fu2003tcp}. However, few of them have been particularly designed for DASH.

\section{Concluding Remarks} \label{sec:conclusion}

We conduct a measurement study on the performance of TCP for chunk delivery in DASH. Our observations and results not only reveal that the chunk request patterns in DASH have a great impact on the performance of TCP strategies, but also identify that conventional TCP strategies may cause user perceived quality degradation in DASH streaming. To improve the streaming quality in DASH according to our measurement studies, we discuss the potential improvement to both DASH request strategies and TCP strategies in slow start and congestion avoidance phases.

\textbf{Increasing CWND in Slow Start.} The problem in the slow resource allocation when small files are delivered in DASH over TCP, is that for most cases the size of CWND is small in the existing TCPs $-$ the size of CWND becomes a bottleneck for chunks transmission. Potential improvement can be summarized as follows.

$\rhd$ First, the intuition for us to improve it is to increase the CWND size to an appropriate value, so as to reduce the time for the CWND size increase at the slow start phase. Such strategies have already been adopted by industrial implementation (e.g., Google TCP improvement).

$\rhd$ Second, especially in DASH, we need to study the impact of chunk size, videos' bitrate, network status and the mobile device status (e.g., the remaining energy), and design a new adaptive CWND optimization scheme taking such information into consideration, instead of just simply mapping chunk size to CWND size. The basic idea is to use cross-layer information in DASH to help CWND fast adapt to an optimal size for later chunks delivery.

\textbf{DASH-Awareness in Congestion Avoidance.} In our measurement studies, we also observe that all of the different types of resource competitions result in degraded quality of streaming in DASH. To enhance the streaming quality, on the other hand, chunk request strategies in DASH also need to be TCP-aware, in a way that a proper time elapse is allowed for TCP to eventually gain resources.

\bibliographystyle{splncs03}
\bibliography{mylib_short}
\end{document}